\documentclass[aps,12pt]{revtex4}
\usepackage{amssymb, amsmath}
\usepackage{graphicx}
\usepackage[latin1]{inputenc}
\usepackage{verbatim}
\def\ltwid{\mathrel{\raise.3ex\hbox{$<$\kern-.75em\lower1ex\hbox{$\sim$}}}}
\def \be{\begin{equation}}
\def \ee{\end{equation}}
\def \bea{\begin{eqnarray}}
\def \eea{\end{eqnarray}}

\def \del{\partial}
\def \a{\alpha}
\def \b{\beta}
\def \f{\frac}

\def \nn{\nonumber}

\begin{document}

\rightline{ITP-UU-80/45, SPIN-08/32} 
\title{Graviton one-loop effective action and inflationary dynamics}
\author{\Large Tomas Janssen\footnote{T.M.Janssen@uu.nl},
 Shun-Pei Miao\footnote{S.P.Miao@phys.uu.nl}
 and Tomislav Prokopec\footnote{T.Prokopec@uu.nl}}

\affiliation{Institute for Theoretical Physics, University of
Utrecht
             Leuvenlaan 4, Postbus 80.195,
              3508 TD Utrecht, The Netherlands}
 \begin{abstract}

We consider the one-loop effective action due to gravitons in a
FLRW background with constant $\epsilon=-{\dot{H}}/{H^2}$. By
expanding around $\epsilon=0$ (corresponding to an expansion
around de Sitter space), we can study how the deviation from de
Sitter space effects the quantum corrected Friedmann equations. We
find that, at zeroth order in $\epsilon$, one-loop effects induce
only a finite shift in the coupling constants. At linear order
in $\epsilon$ there is however a divergent contribution to the equations
of motion. This contribution leads to a nontrivial term in the
renormalized equations that depends logarithmically on $H$ and thus
cannot be absorbed in local counterterms. We find that deviations due
to this term are unobservably small. Our study shows that quantum
effects in quasi de Sitter space can be fundamentally
different then in de Sitter space,
albeit in the case under consideration
the effect is unobservably small.

\end{abstract}




\maketitle

\section{Introduction}
Because of the potential relevance for inflationary cosmology, the
quantum behavior of gravitons on a (locally) de Sitter background
has been a widely studied subject over the past
years~\cite{Allen:1986ta,Antoniadis:1986sb,Allen:1986tt}
\cite{Tsamis:1996qk,Tsamis:1992xa,Tsamis:1996qq}
\cite{Tsamis:1996qm,Higuchi:2000ge,Higuchi:2000ye}
\cite{Finelli:2004bm,Christensen:1979iy,Ford:1984hs}
\cite{Abramo:1996gd,Abramo:1997hu,Abramo:1999wd,Woodard:2004ut}
\cite{Mukhanov:1996ak,Losic:2006ht,Cognola:2005de}
\cite{Abramo:1998hj,Abramo:2001dc,Abramo:2001dd,Iliopoulos:1998wq,Janssen:2007ht}.

One line of research deals with the back-reaction of gravitational
waves on the background
spacetime~\cite{Abramo:1996gd,Abramo:1997hu,Mukhanov:1996ak,Losic:2006ht}.
However of more interest for the present work is the one loop
back-reaction by virtual gravitons on a de Sitter background
which has been calculated by several authors using different
techniques~\cite{Finelli:2004bm}\cite{Tsamis:2005je}\cite{Parker:1969au}.
Since it is not clear whether in these works exactly the same
quantity is calculated and the renormalization schemes differ, the
numerical coefficients differ. However the main result is qualitatively
the same: one loop graviton contributions to the expectation value of
the energy momentum tensor result in a finite, time independent
shift of the effective cosmological constant. Since the
contribution can always be absorbed in a
counterterm~\cite{Tsamis:2005je}, the exact numerical coefficient
has no real physical meaning.\\
The goal of this paper is to go beyond the works mentioned above
and calculate the one loop effective action induced by gravitons
in a more general background space-time using dimensional
regularization. The geometry we consider is a
Friedmann-Lema\^itre-Roberston-Walker (FLRW) geometry with
Hubble parameter $H={\dot{a}}/{a}$ and the additional
constraint that
\begin{equation}\label{eps}
    \epsilon\equiv-\frac{\dot{H}}{H^2}
\end{equation}
is a constant. Standard matter, radiation or dark energy dominated
universes all satisfy this constraint (recall that
in matter era $\epsilon=3/2$, while in radiation era $\epsilon=2$)
and de Sitter space is the special
limit when $\epsilon\rightarrow 0$~\cite{Mukhanov:2005sc}. One
immediate problem with working in such a space-time, instead of 
in de Siter space, is that, for consistency of the Einstein equations,
the addition of matter fields is unavoidable. Whereas in de Sitter
space, the only relevant metric fluctuations are the tensor modes
(gravitational waves), in a more general setting also the scalar
modes, due to the mixing of gravitational and matter degrees of
freedom, have to be taken into
account~\cite{Brandenberger:2003vk}\cite{Mukhanov:1990me}
\cite{Abramo:1998hj,Abramo:2001dc,Abramo:2001dd,Iliopoulos:1998wq}.
This full treatment is considerably more complicated and is
presented elsewhere~\cite{JanssenProkopec:2008}.
For now we will only focus on the
tensor modes, and do not
consider the mixing of degrees of freedom.\\
The main motivation for this work is to show explicitly that new
effects can occur when one considers loop effects in a more
general background then de Sitter space. We find new effects first of
all in quasi de Sitter space, where due to the presence of an
ultraviolet divergence one generates small, but physical
corrections to the quantum Friedmann equations that cannot be
subtracted by local counterterms.

In section~\ref{sgeo} we briefly review our background geometry.
In section~\ref{sprop} we generalize the work
of Ref.~\cite{Janssen:2007ht} and construct the massless minimally
coupled scalar and graviton propagator in
any $\epsilon$=constant space. In section~\ref{soneloop} we
calculate the one-loop effective action contribution to the
quantum corrected Friedmann equations and renormalize the theory
and in section~\ref{sdynamics} we study the associated dynamics
in quasi de Sitter spaces.
We conclude in section~\ref{sconclusion}.
We work in units: $\hbar = 1 = c$.

\section{Geometry}\label{sgeo}

\begin{figure}
\begin{center}
\includegraphics[width=6in]{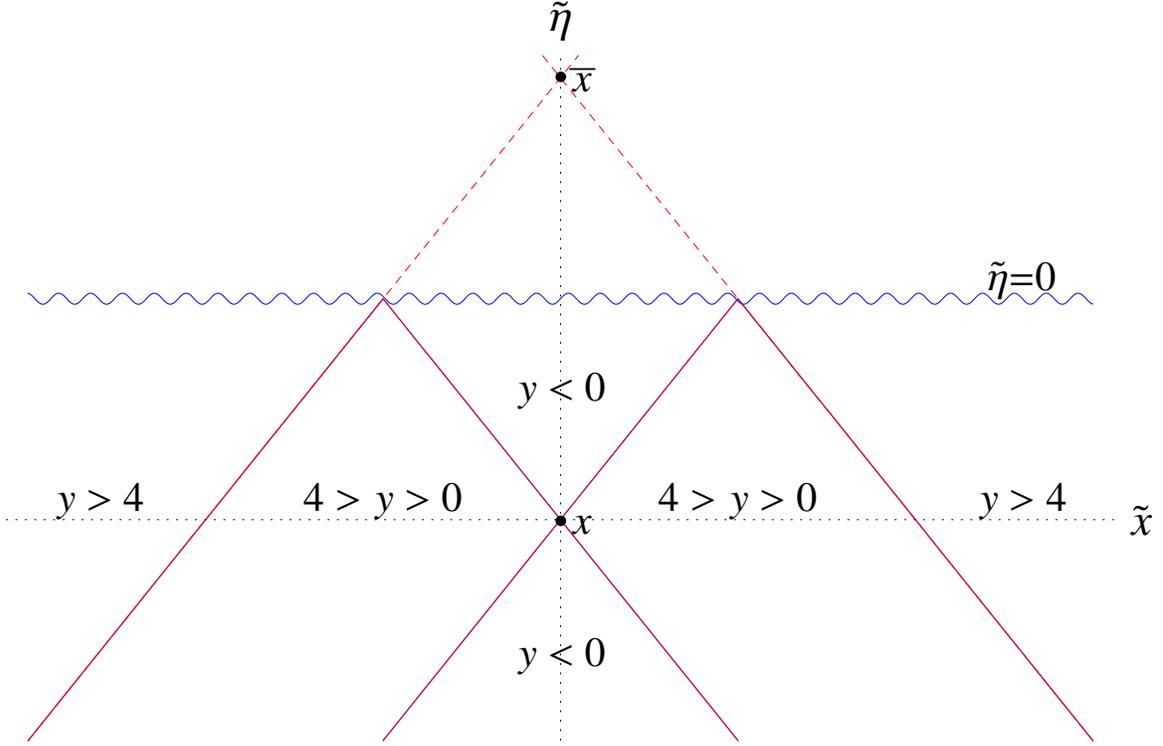}
\caption{The causal structure in the conformal coordinates
(\ref{coord}). The plot assumes $\epsilon<1$, so the coordinates
(of an expanding universe)
cover only the region $\eta<0$. The wavy line at $\eta=0$
indicates future infinity. The lightcone of the point $x$ is given
by $y=0$. If $y=4$, the point $\tilde{x}$ lies on the light cone
of an unobservable image charge at the antipodal point
$\bar{x}$.}\label{spacetime}
\end{center}
\end{figure}

 The geometry we work in is the Friedmann-Lema\^itre-Robertson-Walker
geometry in conformal coordinates
\begin{equation}\label{coord}
    g_{\mu\nu}=a^2\eta_{\mu\nu}\qquad;\qquad\eta_{\mu\nu}=\rm{diag}(-1,1,1,1),
\end{equation}
with the additional constraint,
\begin{equation}\label{epsconstr}
    \epsilon\equiv-\frac{\dot{H}}{H^2}=\text{constant}\qquad;\qquad
    H\equiv \frac{\dot{a}}{a}.
\end{equation}
Here a {\it dot} indicates a derivative with respect to cosmological
time $t$, related to the conformal time $\eta$ by $dt=a d\eta$.
The FLRW geometry obeys the Friedmann equations
\begin{equation}\label{fried}
        \frac{3H^2}{\kappa}-\frac{1}{2}\rho=0\qquad;\qquad
        -\frac{2\dot{H}}{\kappa}-\frac{1}{2}(\rho+p)=0
        \,,\qquad
        \kappa=16\pi G_N
        \,,
\end{equation}
with $G_N$ being the Newton constant, $\rho$ and
$p$ are the energy density and pressure of the cosmological fluid.
If one writes
\begin{equation}
    p=w\rho,
\end{equation}
one immediately finds that (\ref{epsconstr}) implies that $w$ is
constant. One can solve $(\ref{fried})$ for $a$ to find
\begin{equation}\label{adingen}
    \begin{split}
        a(\eta)=\Big((\epsilon-1)H_0\eta\Big)^{-1/(1-\epsilon)}\qquad&;\qquad \epsilon=\frac{3}{2}(1+w)\\
        H=H_0\Big((\epsilon-1)H_0\eta\Big)^{\epsilon/(1-\epsilon)}=H_0 a^{-\epsilon}\qquad&;\qquad  a\eta=-\frac{1}{1-\epsilon}\frac{1}{H}
    \end{split}.
\end{equation}
Notice that if $\epsilon<1$, $\eta$ is negative and if
$\epsilon>1$, $\eta$ is positive. $H_0$ is chosen such that the
$\epsilon\rightarrow 0$ expansion of $H$ corresponds to the one
given in~\cite{Janssen:2007ht}. An important geometrical quantity
is
\begin{equation}
    y\equiv y_{++}=\frac{\Delta x^2_{++}(x;\tilde{x})}{\eta\tilde{\eta}}
    =\frac{1}{\eta\tilde{\eta}}(-(|\eta-\tilde{\eta}|-\imath\varepsilon)^2
    +||\vec{x}-\vec{\tilde{x}}||^2)
\,.
\end{equation}
Here the infinitesimal $\varepsilon>0$ refers to the Feynman
(time-ordered) pole prescription. In de Sitter space $y$ is
related to the geodesic distance $l$ as $y=4\sin^2(\frac{1}{2}H
l)$. If $y<0$, points $\tilde{x}$ are timelike related to $x$, and
if $y>0$, they are spacelike related. We define the antipodal
point $\bar{x}$ of $x$ by the map $\eta\rightarrow -\eta$. Notice
that, since in our coordinates $\eta$ is either positive or
negative, this point is not covered by our coordinates. If $y=4$,
$\tilde{x}$ lies on the lightcone of an unobservable image charge
at the antipodal point $\bar{x}$, see figure \ref{spacetime}.

The curvature tensors are given by
\begin{equation}
    \begin{split}
        R^\alpha{}_{\mu\beta\nu}=&\Big(\frac{a''}{a}-2\Big(\frac{a'}{a}\Big)^2\Big)
        \Big(\delta^\alpha_\nu\delta^0_\mu\delta^0_\beta-\delta^\alpha_0\delta^0_\nu\eta_{\mu\beta}
        -\delta^\alpha_\beta\delta^0_\nu\delta^0_\mu+\delta^0_\beta\delta^\alpha_0\eta_{\mu\nu}\Big)
        \\
        &-\Big(\frac{a'}{a}\Big)^2
        \Big(\delta^\alpha_\nu\eta_{\mu\beta}-\delta^\alpha_\beta\eta_{\mu\nu}\Big)\\
        R_{\mu\nu}=&\Big(\frac{a''}{a}-2\Big(\frac{a'}{a}\Big)^2\Big)\Big(\eta_{\mu\nu}
        -(D-2)\delta_\mu^0\delta_\nu^0\Big)+\Big(\frac{a'}{a}\Big)^2(D-1)\eta_{\mu\nu}\\
        R=&\Big(\frac{a''}{a^3}-2\Big(\frac{a'}{a^2}\Big)^2\Big)2(D-1)
        +\Big(\frac{a'}{a^2}\Big)^2D(D-1)
\,,
    \end{split}
    \label{curvature invariants}
\end{equation}
where $D$ denotes the number of space-time dimensions
and $a^\prime =da/d\eta$.

\section{Scalar Propagator}\label{sprop}

The construction of the graviton propagator in the geometry under
consideration is very similar to the construction in quasi de
Sitter space, as given in~\cite{Janssen:2007ht}. Therefore we will
only give the main steps here.

Since the graviton propagator can
be expressed in terms of massless scalar propagators, we first consider the
following Klein-Gordon equation for a massless scalar in $D$ dimensions
\begin{equation}\label{Scal_KG}
    \sqrt{-g}\left(\Box-\xi
    R\right)\imath\Delta(x;\tilde{x})=\sqrt{-g}\Big[\Box-\xi(D-1)(D-2\epsilon)
    H^2\Big]\imath\Delta(x;\tilde{x})=\imath\delta^D(x-\tilde{x}),
\end{equation}
with some constant $\xi$, where
\begin{equation}\label{Scal_Dalembertian}
  \Box \equiv  \frac{1}{\sqrt{-g}}\partial_\mu \sqrt{-g}g^{\mu\nu}\partial_\nu
         = \frac{1}{a^2}\left(\partial^2 - (D-2)\frac{a'}{a}\partial_0\right)
\,,\qquad (\partial^2 = \eta^{\mu\nu}\partial_\mu\partial_\nu)
\end{equation}
denotes the scalar D'Alembertian.

 We now make the following {\it Ansatz} for the
propagator
\begin{equation}\label{rescale}
\imath\Delta(x;\tilde{x}) =
(a\tilde{a})^{\epsilon(1-D/2)}\imath\Xi(y)
\,.
\end{equation}
After the rescaling (\ref{rescale}), the nonsingular part of the
Klein Gordon equation reads
\begin{equation}\label{scalar_prop_eq_temp2}
    \begin{split}
    &a^D(a\tilde{a})^{\epsilon(1-D/2)}(1-\epsilon)^2H^2
    \Bigg[y(4-y)\big(\frac{d}{dy}\big)^2-\big(D(y-2)\big)\frac{d}{dy}\\
    &\qquad-(1-\epsilon)^{-2}
    \Big((D-1)(D-2\epsilon)\xi-\frac{1}{2}(D-1)(D-2)\epsilon
              +\frac{D}{4}(D-2)\epsilon^2\Big)\Bigg]\imath\Xi(y)=0.
    \end{split}
\end{equation}
This hypergeometric equation has a general solution
\begin{equation}\label{generalhyper}
    \begin{split}
        \imath\Xi(y)&=A\,{}_2F_1\Big(\frac{D-1}{2}+\nu_D,\frac{D-1}{2}-\nu_D;\frac{D}{2};\frac{y}{4}\Big)\\
        &\qquad+B\,{}_2F_1\Big(\frac{D-1}{2}+\nu_D,\frac{D-1}{2}-\nu_D;\frac{D}{2};1-\frac{y}{4}\Big),
    \end{split}
\end{equation}
where
\begin{equation}\label{scalnuorig}
    \nu_D^2=\Big(\frac{D-1}{2}\Big)^2-(1-\epsilon)^{-2}
    \Big[(D-1)(D-2\epsilon)\xi-\frac{1}{2}(D-1)(D-2)\epsilon+\frac{D}{4}(D-2)\epsilon^2\Big].
\end{equation}
The constants $A$ and $B$ are fixed by the singularity conditions.
The vanishing of the antipodal singularity (at $y=4$), which leads
to $\alpha$-vacua
\cite{Einhorn:2002nu,Kaloper:2002cs,Banks:2002nv,Danielsson:2002mb,Goldstein:2003ut},
fixes $A=0$. The constant $B$ is fixed by requiring that the
Hadamard singularity at $y=0$ sources the $\delta$-function
correctly. Notice however that there are values for $\nu_D$ where
this is not possible. In particular if $\nu_D$ is half integer,
larger or equal than 1/2, the hypergeometric equation is no longer a valid
solution. For these particular cases one finds that one cannot
source the $\delta$ function correctly \emph{and} remove the
$\alpha$-vacua. However, our solution is valid arbitrary close to
these points. A well known example of such behavior is the massless
minimally coupled (MMC) scalar field in de
Sitter space~\cite{Allen:1985ux,Mottola:1984ar}. \\
The most singular term sourcing the $\delta$-function is
\begin{equation}
    \imath \Xi(y)_{\rm sing}=B\Big(\frac{y}{4}\Big)^{1-D/2}
    \frac{\Gamma(D/2)\Gamma(D/2-1)}{\Gamma(\frac{D-1}{2}+\nu_D)\Gamma(\frac{D-1}{2}-\nu_D)}
    \,.
\end{equation}
using
\begin{equation}
        \partial^2\frac{1}{\Delta x_{++}^{D-2}}
        =\frac{4\pi^{D/2}}{\Gamma(\frac D2-1)}\imath\delta^D(x-\tilde{x})
\end{equation}
and (\ref{adingen}), we find
\begin{equation}\label{constantB}
    B=\frac{\Gamma(\frac{D-1}{2}+\nu_D)
       \Gamma(\frac{D-1}{2}-\nu_D)}{\Gamma(D/2)}
        \frac{\big(|1-\epsilon|H_0\big)^{D-2}}{(4\pi)^{D/2}}.
\end{equation}
It is important to notice that, due to the rescaling
(\ref{rescale}), $B$ is indeed -- as required -- a constant,
constituting a nontrivial consistency check of
our {\it Ansatz}~(\ref{rescale}).
The MMC {\it scalar propagator}
for a general, constant $\epsilon$ reads
\begin{equation}\label{scal_prop}
    \begin{split}
      \imath\Delta(x;\tilde{x})=&(a\tilde{a})^{-\epsilon(D/2-1)}
      \frac{|1-\epsilon|^{D-2}H_0^{D-2}}{(4\pi)^{D/2}}
      \frac{\Gamma(\frac{D-1}{2}+\nu_D)\Gamma(\frac{D-1}{2}-\nu_D)}{\Gamma(\frac{D}{2})}\\
      &\times{}_2F_1\Big(\frac{D-1}{2}+\nu_D,\frac{D-1}{2}-\nu_D;\frac{D}{2};1-\frac{y}{4}\Big)
      \,.
      \end{split}
\end{equation}
Equation~(\ref{scal_prop}) is the generalization of the scalar
Chernikov-Tagirov propagator~\cite{Chernikov:1968zm} to FLRW spaces
with general, but constant, equation of state parameter $w$, and
thus from Eq.~(\ref{adingen}) a constant $\epsilon$.

\section{Graviton Propagator}\label{gprop}

 Next we consider the graviton propagator. The only difference from
the analysis as given in~\cite{Janssen:2007ht} is in the
coefficients $\nu_{D,n}$. We consider the following action for
gravity plus an arbitrary scalar field $\hat \phi = \hat \phi(x)$.
\begin{equation}
    \begin{split}
    S&=\frac{1}{\kappa}\int d^D x\sqrt{-\hat{g}}\Big(\hat{R}-(D-2)\Lambda\Big)
      + \int d^D x\sqrt{-\hat{g}}
        \Big(
            -\frac{1}{2}\partial_\alpha\hat\phi\partial_\beta\hat\phi
                     \hat{g}^{\alpha\beta}-V(\hat\phi)\Big)\\
    &\equiv S_{\rm{EH}}+S_{\rm{M}}
    \end{split}
\end{equation}
and we write the metric tensor $\hat g$ as a background contribution $g$ plus a
perturbation $h$
\begin{equation}
    \hat{g}_{\mu\nu}=g_{\mu\nu}+h_{\mu\nu}.
\end{equation}
For the scalar fluid $\hat\phi\rightarrow \phi=\phi(t)$,
the background energy density and pressure
are
\begin{eqnarray}
 \rho_M &=& -\frac12(\partial \phi)^2 + V(\phi)
         = \frac12\dot\phi^2 + V(\phi)
\nonumber\\
 p_M &=& -\frac12(\partial \phi)^2 - V(\phi)
      = \frac12\dot\phi^2 - V(\phi)
\,,
\label{rho+p:M}
\end{eqnarray}
where the background scalar field
$\phi = \phi(\eta)$ is a function of (conformal) time only.
The corresponding tree level Friedmann equations are
\begin{equation}\label{background1}
    \begin{split}
    H^2-\f{1}{D-1}\Lambda-\f{\kappa}{(D-1)(D-2)}\rho_M&=0\\
    \dot{H}+\f{D-1}{2}H^2-\f{1}{2}\Lambda
         +\f{\kappa}{2(D-2)}p_M&=0
\\
    \phi''+(D-2)aH\phi'+a^2\f{\partial V}{\partial
    \phi}(\phi)&=0
\,.
    \end{split}
\end{equation}

Expanding the action up to quadratic order in $h_{\mu\nu}$ and using the
background equations of motion for $\phi$
gives~\cite{Christensen:1979iy}\cite{Janssen:2007ht}
\begin{equation}\label{grav_action}
    \begin{split}
    S_{\chi}=\int d^D x&
    \chi_{\alpha\beta}\eta^{\alpha\mu}\eta^{\beta\nu}
       \Bigg[\Big(\partial^2+\frac{D-2}{2}\dot{H}a^2+\frac{D}{4}(D-2)H^2
    a^2\Big)\Big(\frac{1}{4}\delta^\rho_\mu\delta^\sigma_\nu-\frac{1}{8}\eta_{\mu\nu}\eta^{\rho\sigma}\Big)\\
    &\hskip 2cm
 -\frac{D-2}{2}(H^2+\dot{H})a^2\delta^0_\mu\delta^\rho_\nu\delta^\sigma_0\Bigg]
   \chi_{\rho\sigma}
 + S_{\rm os}
   \,,
    \end{split}
\end{equation}
where we defined,
\begin{equation}
    h_{\mu\nu}=\sqrt{\kappa}a^{2}\psi_{\mu\nu}
              = \sqrt{\kappa}a^{3-D/2}\chi_{\mu\nu},
    \quad\tilde{\chi}_{\mu\nu}=\chi_{\mu\nu}
    -\frac{1}{2}g_{\mu\nu}\chi,\quad\chi=g^{\mu\nu}\chi_{\mu\nu}
\,,\quad (\psi_{\mu\nu}= a^{1-D/2}\chi_{\mu\nu})
\,,
\end{equation}
we added a gauge fixing term
\begin{equation}\label{gaugefix}
    -\frac{1}{2}a^{D+4}\Big[\nabla_\mu(a^{1-D/2}\tilde{\chi}^{\mu\alpha})\Big]
    \nabla_\nu(a^{1-D/2}\tilde{\chi}^\nu{}_\alpha)
\end{equation}
and where $S_{\rm os}$ denotes the second-order terms that vanish on-shell
({\it cf.} Eqs.~(\ref{background1})),
\begin{eqnarray}
 S_{\rm os}\! &=&\! -\! \int \! d^D x a^2\chi_{\alpha\beta}
   \left(\frac14\eta^{\rho\alpha}\eta^{\sigma\beta}
        - \frac18 \eta^{\alpha\beta}\eta^{\rho\sigma}\right)
      2(D\!-\!2)\Big(\dot H + \frac{D\!-\!1}{2}H^2 - \frac{\Lambda}{2}
                    + \frac{\kappa p_M}{2(D\!-\!2)}\Big)\chi_{\rho\sigma}
\nonumber\\
 &&+\, \int d^D x a^2\chi_{\alpha\beta}
   \left(-\delta^{(\alpha}_0\eta^{\beta)(\rho}\delta^{\sigma)}_0
         + \frac12\delta^{(\alpha}_0\delta^{\beta)}_0\eta^{\rho\sigma}
   \right)
     \Big((D\!-\!2)\dot H+\frac{\kappa}{2}(\rho_M+p_M)\Big)\chi_{\rho\sigma}
\,.
\label{Sos}
\end{eqnarray}
Even though these terms do not contribute to the graviton propagator,
they do contribute to the one loop effective action
and hence we must keep them.

From the action (\ref{grav_action}) we find the graviton
propagator for $\chi_{\mu\nu}$
\begin{equation}\label{gravprop}
    a^{1-D/2}\tilde{a}^{1-D/2}\imath[_{\rho\sigma}\Delta^{\alpha\beta}]
 =(T_0)_{\rho\sigma}^{\quad\alpha\beta}
  \imath\Delta_0
+(T_1)_{\rho\sigma}^{\quad\alpha\beta}\imath\Delta_1
  +(T_0)_{\rho\sigma}^{\quad\alpha\beta}\imath\Delta_2,
\end{equation}
where
\begin{eqnarray}
    (T_0)_{\rho\sigma}^{\quad\alpha\beta}
 &=& 2\bar{\delta}^{(\alpha}_\rho\bar{\delta}^{\beta)}_
 \sigma-\frac{2}{D-3}\bar{\eta}_{\rho\sigma}\bar{\eta}^{\alpha\beta}
\,,\quad
(T_1)_{\rho\sigma}^{\quad\alpha\beta}
=4\delta_{(\rho}^0\bar{\delta}_{\sigma)}^{(\alpha}\delta^{\beta)}_0
\nonumber\\
 (T_2)_{\rho\sigma}^{\quad\alpha\beta}
 &=& \frac{2}{(D-2)(D-3)}(\eta_{\rho\sigma}
 +(D-2)\delta_\sigma^0 \delta_\rho^0)(\eta^{\alpha\beta}
 +(D-2)\delta^\beta_0\delta^\alpha_0)
\label{TensorStructures}
 \,
\end{eqnarray}
denote the relevant graviton tensor structures and
\begin{equation}
    \bar{\eta}_{\mu\nu}=\eta_{\mu\nu}+\delta_\mu^0 \delta_\nu ^0
\end{equation}
is the spatial part of the metric tensor.
The prefactor $a^{1-D/2}\tilde{a}^{1-D/2}$ in Eq.~(\ref{gravprop})
comes from the fact that the $\chi_{\mu\nu}$ field
differs by a rescaling, $\chi_{\mu\nu}=a^{-1+D/2}\psi_{\mu\nu}$,
from the $\psi_{\mu\nu}$ field for which the propagator is
calculated in~\cite{Janssen:2007ht}. The scalar propagators
$\Delta_n$ for the pseudograviton $\psi_{\mu\nu}$ obey
\begin{equation}\label{scalarprops}
  \sqrt{-g}\Big(\Box-n(D-n-1)(1-\epsilon)H^2\Big)\imath\Delta_n
  = \imath\delta^D(x-x')\,,
\qquad (n=0,1,2)
\,,
\end{equation}
and thus they are given by (\ref{scal_prop}) with
\begin{equation}\label{nugrav}
  \nu_{D,n}^2 = \frac{(D-1)^2}{4}
  +\frac{\frac{1}{2}(D-1)(D-2)\epsilon
         - n(1-\epsilon)(D-n-1)
         -\frac{D}{4}(D-2)\epsilon^2}
        {(1-\epsilon)^{2}}.
\end{equation}
The ghost action associated with our gauge fixing is given by
\begin{equation}\label{ghostlagfinal}
    S_{\rm ghost}=\int d^D x\eta_{\alpha\beta}
    \bar{U}^\beta\Big[\Big(\partial^2
                 +\frac{D-2}{2}\dot{H}a^2+\frac{D}{4}(D-2)H^2
    a^2\Big)\delta^\alpha_\mu-(D-2)a^2(H^2-\dot{H})\delta_\mu^0\delta_0^\alpha\Big]U^\mu,
\end{equation}
where $U$ is the ghost field, related to the ghost field of
Ref.~\cite{Janssen:2007ht} by a scale transformation $U=a^{D/2-1}V$.
The ghost propagator is found to be
\begin{equation}
    a^{1-D/2}\tilde{a}^{1-D/2}\imath[_\alpha\hat{\Delta}^\rho](x;\tilde{x})=\imath\bar{\delta}_\alpha^\rho\hat{\Delta}_0(x;\tilde{x})+\imath\delta_\alpha^0\delta^\rho_0\hat{\Delta}_1(x;\tilde{x}),
\end{equation}
where the $\hat{\Delta}_n$ propagators satisfy
\begin{equation}\label{ghostprops}
  \sqrt{-g}\Big(\Box-n(D-n-1)(1+\epsilon)H^2\Big)\imath\hat{\Delta}_n
  = \imath\delta^D(x-x')\,, \qquad (n=0,1)
\,.
\end{equation}
Therefore the corresponding propagators are given by~(\ref{scal_prop}) with
\footnote{The analysis of Ref.~\cite{JanssenProkopec:2008} shows that,
when the mode graviton-matter mixing is taken account of,
the ghost propagators
$\imath\hat{\Delta}_n(x;\tilde{x})$ ($n=0,1$) become identical to the
graviton propagators $\imath{\Delta}_n(x;\tilde{x})$ ($n=0,1$).
}
\begin{equation}\label{nughost}
  \hat{\nu}_{D,n}^2
  = \frac{(D-1)^2}{4}
  +\frac{\frac{1}{2}(D-1)(D-2)\epsilon
         - n(1+\epsilon)(D-n-1)
         - \frac{D}{4}(D-2)\epsilon^2}
        {(1-\epsilon)^{2}}
\,.
\end{equation}
Note that both the graviton and ghost propagators given above differ
by a scaling $a^{1-D/2}\tilde{a}^{1-D/2}$ from the ones given
in Ref.~\cite{Janssen:2007ht}.

\section{One-loop effective action}\label{soneloop}

 The one-loop effective action is defined as~\cite{Birrell:1982ix}
\begin{equation}
    \Gamma=-\imath\langle\rm{out},0|0,\rm{in}\rangle.
\end{equation}
While in flat space and in the absence of external sources
such a vacuum-to-vacuum transition
can be normalized to unity, this is not possible in
general curved space-times. Since our lagrangian is
quadratic in the graviton and ghost fields, we can integrate them
out to get the one loop effective action
\begin{equation}\label{effective1}
    \begin{split}
    \exp[\imath \Gamma]&=\int\mathcal{D}h_{\mu\nu}\mathcal{D}
    W\mathcal{D}\bar{W}
    \exp\Big[\imath\big(S_{\rm{EH}}+S_{\rm{M}}+S_\chi + S_{\rm os}
                        + S_{\rm ghost}\big)
     \Big]\\
    &=\int\mathcal{D}\chi_{\mu\nu}\mathcal{D}U\mathcal{D}\bar{U}
        \exp\Big[\imath\big(S_{\rm{EH}}+S_{\rm{M}}+S_\chi + S_{\rm os}
                +S_{\rm ghost}\big)\Big]\\
    &=\exp\Big[\imath\big(S_{\rm{EH}}+S_{\rm{M}}\big)\Big]
       \frac{\det(\mathcal{F^\alpha{}_\mu})}
            {\sqrt{\det(\mathcal{D^{\rho\sigma}{}_{\mu\nu}}
                    + \mathcal{\delta D^{\rho\sigma}{}_{\mu\nu}})}}
\,,
    \end{split}
\end{equation}
where the step from the first to the second line can be made by
noticing that the Jacobian of the transformation contributes as a
D-dimensional delta function evaluated at zero, $\delta^D(0)$. In dimensional
regularization such a term does not contribute. In
(\ref{effective1}) $\mathcal{D}$ and $\mathcal{F}$ are the kinetic
operators from the rescaled graviton~(\ref{grav_action}),
including the off-shell contribution~(\ref{Sos}),
and the ghost~(\ref{ghostlagfinal}), respectively, given by
\begin{eqnarray}
    \mathcal{D}^{\rho\sigma}{}_{\mu\nu}
&=&\Big(\partial^2+\frac{1}{4}(D-2)(D-2\epsilon)H^2
    a^2\Big)\Big(\frac{1}{2}\delta^\rho_\mu\delta^\sigma_\nu
      -\frac{1}{4}\eta_{\mu\nu}\eta^{\rho\sigma}\Big)
 -(D-2)(1-\epsilon)H^2a^2\delta^0_{(\mu}\delta^{(\rho}_{\nu)}\delta^{\sigma)}_0
\nonumber\\
\delta D^{\rho\sigma}{}_{\mu\nu}
&=& - a^2
   \left(\frac12\delta^{(\rho}_\mu\delta^{\sigma)}_\nu
        - \frac14 \eta^{\rho\sigma}\eta_{\mu\nu}\right)
      2(D\!-\!2)\Big(\dot H + \frac{D\!-\!1}{2}H^2 - \frac{\Lambda}{2}
                    + \frac{\kappa p_M}{2(D\!-\!2}\Big)
\nonumber\\
 &&+\, a^2
   \left(2\delta^{(\rho}_0\delta^{\sigma)}_{(\mu}\delta_{\nu)}^0
         + \delta^{(\rho}_0\delta^{\sigma)}_0\eta_{\mu\nu}
   \right)
     \Big((D\!-\!2)\dot H+\frac{\kappa}{2}(\rho_M+p_M)\Big)
\nonumber\\
  \mathcal{F}^\alpha{}_\mu
  &=& \Big(\partial^2+\frac14(D-2)(D-2\epsilon)H^2
 a^2\Big)\delta^\alpha_\mu-(D-2)(1+\epsilon)H^2a^2\delta_\mu^0\delta^\alpha_0
\,.
\end{eqnarray}
From Eq.~(\ref{effective1}) we obtain
\begin{equation}\label{effective5}
    \begin{split}
    \Gamma&=S_{\rm{EH}}+S_{\rm{M}}
     +\frac{\imath}{2}\rm Tr
        \ln[\mathcal{D^{\rho\sigma}{}_{\mu\nu}}
            +\mathcal{\delta D^{\rho\sigma}{}_{\mu\nu}}]
     -\imath\rm Tr\ln[\mathcal{F^\alpha{}_\mu}]\\
    &\equiv S_{\rm{EH}}+S_{\rm{M}}+\Gamma_{1L}
\,.
    \end{split}
\end{equation}
While in principle one could -- at least formally -- evaluate the
effective action, the object one is eventually interested in is
the effective Friedmann equation, {\it i.e.} the equations of motion of
the metric. Moreover in the present case there is the technical
complication that we need to work under the constraint that
$\epsilon$ is constant.
As long as $\dot \epsilon$ remains small,
there is no problem with imposing
such a constraint in the equations of motion.
On the other hand, imposing such a constraint in
the action might change the dynamics substantially.
By taking the functional derivative with respect
to the scale factor $a=a(\eta)$,
we obtain the Einstein trace equation, that is the $-(00)+3(ii)$
component of the Einstein equation.
 Since in a FLRW universe there are only two independent
equations, the second equation can be obtained by imposing the
Bianchi identity. Thus our first equation of motion is given by
\begin{equation}\label{effect}
    \begin{split}
        \frac{\delta \Gamma}{\delta a(l)}&=\frac{\delta (S_{\rm
        HE}+ S_{\rm M})}{\delta a(l)}+\frac{\delta \Gamma_{1\rm
        L}}{\delta a(l)}\\
        &=Va^3\bigg[\frac{24}{\kappa}
          \Big(H^2-\frac13\Lambda+\frac12\dot{H}\Big)
         + 3p_M-\rho_M\bigg]
        +\frac{\delta \Gamma_{1\rm L}}{\delta a(l)}
        \,,
    \end{split}
\end{equation}
where $V=\int d^{D-1}x$ denotes the volume of space and
$p_M$ and $\rho_M$ are the pressure and energy density
associated to the matter action $S_{\rm M}$ and they are defined by,
$T_{\mu\nu}^M = (2/\sqrt{-g})\delta S_M/\delta g^{\mu\nu}
= - g_{\mu\nu}p_M - a^2\delta_\mu^{\;0}\delta_\nu^{\;0}(\rho_M+p_M)$.
We first focus on the
graviton contribution to $\delta \Gamma_{1\rm L}/\delta a(l)$.
We first write the effective graviton action in terms of $a$:
\begin{eqnarray}
\Gamma_{\rm g}[a]
 &=&\f{\imath}{2}\,\textrm{Tr}\,\ln\,\biggl\{
 \Bigl[\eta_{0(\mu}\eta_{\nu)(\rho}\eta_{\sigma)0}\Bigr]
(D-2)\Bigl(\f{a''}{a}-\f{{a'}^{2}}{a^2}\Bigr)
\nn\\
&+&\Bigl[\,\f{1}{2}\eta_{\mu(\rho}\eta_{\sigma)\nu}
-\f{1}{4}\eta_{\mu\nu}\eta_{\rho\sigma}\Bigr]
\biggl(\del^2+\f{1}{2}(D-2)
\Bigl[\f{1}{2}(D-4)\f{{a'}^{2}}{a^2}+\f{a''}{a}\Bigr]\biggr)
 + \mathcal{\delta D_{\rho\sigma\,\mu\nu}}
\biggr\}\,.
\label{dG/da:1}
\end{eqnarray}
Now, since the term within the logarithm is just the kinetic operator,
upon variation we will generate the inverse of this object. This
inverse is of course the propagator. Taking the trace implies here
both tracing over the indices, and evaluation at
coincidence~\cite{Birrell:1982ix}.
After taking the functional
derivative we get,
\begin{eqnarray}
\frac{1}{V}\f{\delta\Gamma_{g}[a]}{\delta a(l)}
&=&
\frac{1}{V}\f{\delta\Gamma'_{g}[a]}{\delta a(l)}
+\frac{1}{V}\f{\delta\Gamma''_{g}[a]}{\delta a(l)}
\end{eqnarray}
where the latter term originates from variation of
$\mathcal{\delta D_{\rho\sigma\,\mu\nu}}$ in Eq.~(\ref{dG/da:1}).
We have
\begin{eqnarray}
\frac{1}{V}\f{\delta\Gamma'_{g}[a]}{\delta a(l)}
 &=&(D-2)\biggl\{
\Bigl(-\delta^0_{(\mu}\delta_{\nu)}^{(\rho}
\delta^{\sigma)}_{0}\Bigr)\f{1}{2a}\f{d^2}{dl^2}
+\Bigl(\f{1}{2}\delta_{\mu}^{(\rho} \delta^{\sigma)}_{\nu}
\!-\!\f{1}{4}\eta_{\mu\nu} \eta^{\rho\sigma}\Bigr)
\label{dGamma'}
\\
&&\hspace{1.3cm}\times\,
\biggl[\frac{D\!-\!2}{4}
\Bigl(\f{a^{'2}}{a^3}-\f{a''}{a^2}
-\f{a'}{a^2}\f{d}{dl}\Bigr)+\f{1}{4a}\f{d^2}{dl^2}
\,\biggr]
\,\biggr\}\,\eta^{\mu(\a}\eta^{\b)\nu}\,
\imath\Bigl[\mbox{}_{\a\b}\Delta_{\rho\sigma}\Bigr](x;x)
\,,\quad
\nn
\end{eqnarray}
and
\begin{eqnarray}
&&\frac{1}{V}\f{\delta\Gamma''_{g}[a]}{\delta a(l)}
=(D-2)\biggl\{
 \Bigl(\f{1}{2}\delta_{\mu}^{(\rho} \delta^{\sigma)}_{\nu}
-\f{1}{4}\eta_{\mu\nu} \eta^{\rho\sigma}\Bigr)
\label{dgammada}
\\
 &&\hspace{4.cm}
  \times\, \bigg[(D\!-\!3)
           \bigg(
               \frac{a''}{a^2}-\frac{{a'}^2}{a^3}
               + \frac{a'}{a^2}\frac{d}{dl}
            \bigg)
               + a\Lambda + a\frac{\kappa}{D\!-\!2}V(\phi)
               - \frac{1}{a}\frac{d^2}{dl^2}
    \bigg]
\nn\\
&&\hspace{0.4cm}
+\,\Bigl(\delta^0_{(\mu}\delta_{\nu)}^{(\rho} \delta^{\sigma)}_0
+\f{1}{2}\delta^0_{(\mu}\delta_{\nu)}^0 \eta^{\rho\sigma}\Bigr)
    \bigg[2\bigg(
               \frac{a''}{a^2}-\frac{{a'}^{\,2}}{a^3}
               + \frac{a'}{a^2}\frac{d}{dl}
            \bigg)
               + \frac{1}{a}\frac{d^2}{dl^2}
    \bigg]
\,\biggr\}\,\eta^{\mu(\a}\eta^{\b)\nu}\,
\imath\Bigl[\mbox{}_{\a\b}\Delta_{\rho\sigma}\Bigr](x;x)
\,,\quad
\nn
\end{eqnarray}
where $\Bigl[\mbox{}_{\a\b}\Delta_{\rho\sigma}\Bigr](x;x)$ denotes
the graviton propagator (\ref{gravprop}) evaluated at coincidence.
Eq.~(\ref{TensorStructures}) implies the following contractions,
\begin{eqnarray}
\eta^{\mu(\a}\eta^{\b)\nu}\Bigl(\f{1}{2}\delta_{\mu}^{(\rho}
\delta^{\sigma)}_{\nu} -\f{1}{4}\eta_{\mu\nu}
\eta^{\rho\sigma}\Bigr)(T_{0})_{\a\b\rho\sigma}
  &=&\f{1}{2}D(D\!-\!1)
\,,\nn\\
\eta^{\mu(\a}\eta^{\b)\nu}\Bigl(\f{1}{2}\delta_{\mu}^{(\rho}
\delta^{\sigma)}_{\nu} -\f{1}{4}\eta_{\mu\nu}
\eta^{\rho\sigma}\Bigr)(T_{1})_{\a\b\rho\sigma}
  &=&D\!-\!1
\,,\nn\\
\eta^{\mu(\a}\eta^{\b)\nu}\Bigl(\f{1}{2}\delta_{\mu}^{(\rho}
\delta^{\sigma)}_{\nu} -\f{1}{4}\eta_{\mu\nu}
\eta^{\rho\sigma}\Bigr)(T_{2})_{\a\b\rho\sigma}
  &=&1
\,,\nn\\
\eta^{\mu(\a}\eta^{\b)\nu}\Bigl(\delta^0_{(\mu}\delta_{\nu)}^{(\rho}
\delta^{\sigma)}_{0}\Bigr)(T_1)_{\a\b\rho\sigma}
  &=&D\!-\!1
\,,\nn\\
\eta^{\mu(\a}\eta^{\b)\nu}\Bigl(\delta^0_{(\mu}\delta_{\nu)}^{(\rho}
\delta^{\sigma)}_{0}\Bigr)(T_2)_{\a\b\rho\sigma}
  &=&\f{2(D\!-\!3)}{D\!-\!2}
\,,\nn\\
\eta^{\mu(\a}\eta^{\b)\nu}\Bigl(\delta^0_{(\mu}\delta_{\nu)}^0
\eta^{\rho\sigma} \Bigr)(T_2)_{\a\b\rho\sigma}
  &=&\f{4}{D\!-\!2}
\,,
\label{contraction}
\end{eqnarray}
where $(T_{i})_{\a\b\rho\sigma} \equiv
 (T_{i})_{\a\b}^{\quad\gamma\delta}\eta_{\gamma\rho}\eta_{\delta\sigma}$
($i=0,1,2$). Other contractions vanish.
After substituting (\ref{contraction}) into (\ref{dgammada}) one
obtains
\begin{eqnarray}
&&\hspace{-0.7cm}\frac{1}{V}\f{\delta\Gamma'_{g}[a]}{\delta
a(l)}=\f{1}{8}D(D-1)(D-2)
\biggl[\,\Bigl(D-2\Bigr)\Bigl(\f{{a'}^2}{a^3}-\f{a''}{a^2}
-\f{a'}{a^2}\f{d}{dl}\Bigr)
+\f{1}{a}\f{d^2}{dl^2}\,\biggr]\imath\Delta_{0}(x;x)a^{D-2}
\nn\\
&&\hspace{1.5cm}+\f{1}{4}(D-1)(D-2)
\biggl[\,\Bigl(D-2\Bigr)\Bigl(\f{{a'}^2}{a^3}-\f{a''}{a^2}
-\f{a'}{a^2}\f{d}{dl}\Bigr)
-\f{1}{a}\f{d^2}{dl^2}\,\biggr]\imath\Delta_{1}(x;x)a^{D-2}
\nn\\
&&\hspace{1.5cm}+\biggl[\,\f{1}{4}(D-2)^2\Bigl(\f{{a'}^2}{a^3}-\f{a''}{a^2}
-\f{a'}{a^2}\f{d}{dl}\Bigr)-\f{1}{4}\Bigl(3D-10\Bigr)
\f{1}{a}\f{d^2}{dl^2}\,\biggr]\imath\Delta_{2}(x;x)a^{D-2}
\,.\quad
\label{dGamma':a}
\end{eqnarray}
and
\begin{eqnarray}
\frac{1}{V}\f{\delta\Gamma''_{g}[a]}{\delta a(l)}
&=&(D-2)
\biggl\{\frac{D(D\!-\!1)}{2}\bigg[
              \bigl(D\!-\!3\bigr)\Bigl(\f{a''}{a^2}-\f{{a'}^2}{a^3}
                    +\f{a'}{a^2}\f{d}{dl}\Bigr)
                     + a\Lambda + a \frac{\kappa V(\phi)}{D\!-\!2}
                     -\f{1}{a}\f{d^2}{dl^2}\,\bigg]
\nn\\
&&\hskip 3.5cm   \times\, \imath\Delta_{0}(x;x)a^{D-2}
\label{dGamma''}
\\
&+&(D\!-\!1)\bigg[(D\!-\!1)\Bigl(\f{a''}{a^2}-\f{{a'}^{2}}{a^3}
+\f{a'}{a^2}\f{d}{dl}\Bigr)
 + a\Lambda + a \frac{\kappa V(\phi)}{D\!-\!2}
\bigg]\imath\Delta_{1}(x;x)a^{D-2}
\nn\\
&+&\bigg[(D\!+\!1)\Bigl(\f{a''}{a^2}-\f{{a'}^{2}}{a^3}
+\f{a'}{a^2}\f{d}{dl}\Bigr)
 + a\Lambda + a \frac{\kappa V(\phi)}{D\!-\!2}
 +\f{1}{a}\f{d^2}{dl^2}\,\bigg]\imath\Delta_{2}(x;x)a^{D-2}
\biggr\}
\,.\quad
\nn
\end{eqnarray}
This can be further simplified by making use of the on-shell
relation~(\ref{background1})
\begin{equation}
 \Lambda + \frac{\kappa V(\phi)}{D\!-\!2}
  = (D\!-\!1) H^2 + \dot H
\,.
\label{os:3}
\end{equation}

\medskip

For the ghost field we follow exactly the same procedure to obtain
\begin{eqnarray}
\hspace{-0.1cm}\frac{1}{V}\f{\delta\Gamma_{gh}[a]}{\delta a(l)}
&=&(D-2)\biggl\{\,-\f{1}{2}(D-1)\biggl[(D-2)
\Bigl(\f{{a'}^{\,2}}{a^3}-\f{a''}{a^2}-\f{a'}{a^2}\f{d}{dl}\Bigr)
+\f{1}{a}\f{d^2}{dl^2}\,\biggr]\imath\hat{\Delta}_0(x;x)a^{D-2}
\nn\\
\hspace{0.cm}&+&\!\biggl[-\f{1}{2}(D-10)
\Bigl(\f{{a'}^{\,2}}{a^3}-\f{a''}{a^2}-\f{a'}{a^2}\f{d}{dl}\Bigr)
-\f{3}{2}\f{1}{a}\f{d^2}{dl^2}\,\biggr]\imath\hat{\Delta}_1(x;x)a^{D-2}
  \biggr\}
\,.
\label{dgammada1}
\end{eqnarray}
Upon combining (\ref{dGamma':a}) and (\ref{dgammada1}) we obtain,
\begin{eqnarray}
\hspace{-0.1cm}
\frac{1}{V}\f{\delta\Gamma'_{g+gh}[a]}{\delta
a(l)}\!&=&\!(D\!-\!2)
\biggl\{\,\f{1}{8}(D\!-\!1)(D\!-\!4)\biggl[(D\!-\!2)
\Bigl(\f{{a'}^{2}}{a^3}\!-\!\f{a''}{a^2}\!-\!\f{a'}{a^2}\f{d}{dl}\Bigr)
\!+\!\f{1}{a}\f{d^2}{dl^2}\,\biggr]\imath\Delta_0(x;x)a^{D-2}
\nn\\
&+&\f{1}{4}(D-1)\biggl[(D-2)
\Bigl(\f{{a'}^{2}}{a^3}-\f{a''}{a^2}-\f{a'}{a^2}\f{d}{dl}\Bigr)
-\f{1}{a}\f{d^2}{dl^2}\,\biggr]\,\imath\Delta_1(x;x)a^{D-2}
\nn\\
&+&\biggl[\,-\f{1}{2}(D-10)
\Bigl(\f{{a'}^{2}}{a^3}-\f{a''}{a^2}-\f{a'}{a^2}\f{d}{dl}\Bigr)
-\f{3}{2}\f{1}{a}\f{d^2}{dl^2}\,\biggr]\,\imath\hat{\Delta}_1(x;x)a^{D-2}
\nn\\
&+&\biggl[\,\f{1}{4}(D-2)
\Bigl(\f{{a'}^{2}}{a^3}-\f{a''}{a^2}-\f{a'}{a^2}\f{d}{dl}\Bigr)
-\f{1}{4}\f{(3D-10)}{(D-2)}\f{1}{a}\f{d^2}{dl^2}\,\biggr]
\,\imath\Delta_2(x;x)a^{D-2}\,\biggr\}
\,,
\label{dgammada2}
\end{eqnarray}
where we made use of $\imath \Delta_0 = \imath\hat\Delta_0$.

The next step is to evaluate the propagator at coincidence. From
this point on we need to constrain our calculation to the case
where $\epsilon$ is constant. In this case the coincidence limit
of the propagators are given by
\begin{eqnarray}
\hspace{-.cm}\imath\Delta_{n}(x;x)
&=& |1-\epsilon|^{D-2}
H^{D-2}\f{\Gamma(1-\f{D}{2})}{(4\pi)^{\f{D}{2}}}
\f{\Gamma(\f{D-1}{2}+\nu_{D,\,n})\Gamma(\f{D-1}{2}-\nu_{D,\,n})}
{\Gamma(\f{1}{2}+\nu_{D,\,n})\Gamma(\f{1}{2}-\nu_{D,\,n})}
\,
\label{propagator}\\
\hspace{-.cm}\f{d}{d\eta}a^{D-2}\imath\Delta_{n}(x;x)
&=&Ha(D-2)(1-\epsilon)a^{D-2}\imath\Delta_{n}(x;x)
\,
\label{dpropagator}\\
\hspace{-.cm}\f{d^2}{d\eta^2}a^{D-2}\imath\Delta_{n}(x;x)
&=&H^{2}a^{2}
(D-1)(D-2)(1-\epsilon)^2\,a^{D-2}\imath\Delta_{n}(x;x)
\,,
\label{ddpropagator}
\end{eqnarray}
where the $\nu$ parameters for the ghost and the graviton are given
by (\ref{nughost}) and (\ref{nugrav}), respectively.
Notice that the propagator
is \emph{first} evaluated at coincidence and only then hit by the
derivative. Using these results in Eqs.~(\ref{dgammada2})
and~(\ref{dGamma''}) and Eq.~(\ref{os:3}) we obtain
\begin{eqnarray}
\frac{1}{V}\f{\delta\Gamma'_{g+gh}[a]}{\delta a(l)}
&=&(D-1)(D-2)(1-\epsilon)H^{2}a^{D-1}
\biggl\{-\f{1}{8}(D-1)(D-2)(D-4)
\epsilon\,\imath\Delta_0(x;x)
\nn\\
&&\hspace{5.5cm}-\,\f{1}{4}(D-1)(D-2)
(2-\epsilon)\,\imath\Delta_{1}(x;x)
\label{simplifypp}
\\
&&\hspace{5.5cm}-\,\f{1}{2}\Bigl[\,2(D+2)-3(D-2)\epsilon\,\Bigr]\,
\imath\hat{\Delta}_{1}(x;x)\nn\\
&&\hspace{5.5cm}-\,\f{1}{4}\Bigl[\,4(D-3)-(3D-10)\epsilon\,\Bigr]\,
\imath\Delta_{2}(x;x)\biggr\}
\,
\nonumber
\end{eqnarray}
and
\begin{eqnarray}
\frac{1}{V}\f{\delta\Gamma''_{g}[a]}{\delta a(l)}
&=&(D-2)H^2 a^{D-1}
\biggl\{\frac{D(D\!-\!1)(D\!-\!2)\epsilon}{2}\Big[D-(D\!-\!1)\epsilon\Big]
                       \imath\Delta_{0}(x;x)
\nn\\
&&+\,(D\!-\!1)\Big[D(D\!-\!1)-(D^2\!-\!2D\!+\!2)\epsilon\Big]
             \imath\Delta_{1}(x;x)
\nn\\
&&+\,\Big[2D(D\!-\!1)-(3D^2\!-\!6D\!+\!4)\epsilon
       + (D\!-\!1)(D\!-\!2)\epsilon^2
   \Big]\imath\Delta_{2}(x;x)
\biggr\}
\,.\quad
\label{dGamma'':b}
\end{eqnarray}

\bigskip

We substitute (\ref{propagator}) in (\ref{simplifypp})
and~(\ref{dGamma'':b}), add the two contributions
and expand around $D=4$ to obtain the nonrenormalized one loop effective
action
\footnote{
Had we not included the contribution~(\ref{dGamma'':b}) 
from the off shell
terms~(\ref{Sos}) in the effective action,
instead of Eq.~(\ref{un-renormalized})) we would get:
\[
       \frac{1}{H^{4}a^{D-1}V} \f{\delta\Gamma'_{g+gh}[a]}{\delta a(l)}
=
 -\frac{21\mu^{D-4}\epsilon(1-\epsilon)(2-\epsilon)}{4\pi^2(D-4)}
+\frac{1-\epsilon}{16\pi^2}\bigg\{\Big(-12+22\epsilon-35\epsilon^2+30\epsilon^3\Big)
\]
\[
       +6\epsilon(2-\epsilon)
  \bigg[-7\ln\Big[\frac{(1-\epsilon)^2}{4\pi}\Big]
        -14\ln\Big(\frac{H}{\mu}\Big)
        - 7 \gamma_E
        + 2\psi\Big(\frac{1}{1-\epsilon}\Big)
        + 2\psi\Big(-\frac{\epsilon}{1-\epsilon}\Big)
\]
\[
\hskip 1.5cm
 -\,9\bigg(
     \psi\Big(\frac12+\frac{\sqrt{1-14\epsilon+\epsilon^2}}{2(1-\epsilon)}\Big)
   +\psi\Big(\frac12-\frac{\sqrt{1-14\epsilon+\epsilon^2}}{2(1-\epsilon)}\Big)
     \bigg)
       \bigg]\bigg\}
+\mathcal{O}(D-4)
\,.
\]
Note that the structure of the divergent term in this expression
is simpler such that -- unlike the divergence in Eq.~(\ref{un-renormalized})
-- it can be renormalized by making use of the $R^2$ counterterm alone.
}
\begin{eqnarray}
       \frac{1}{H^{4}a^{D-1}V} \f{\delta\Gamma_{g+gh}[a]}{\delta a(l)}
&=&
 -\frac{\epsilon(198-241\epsilon+63\epsilon^2)}{4\pi^2}
       \frac{\mu^{D-4}}{D\!-\!4}
\nn\\
&+&\frac{1}{16\pi^2}\bigg\{\Big(84-1810\epsilon+2307\epsilon^2
                              -791\epsilon^3+54\epsilon^4\Big)
\nonumber\\
       &&-\,2\epsilon(198-241\epsilon+63\epsilon^2)
  \bigg[\ln\Big(\frac{(1-\epsilon)^2}{4\pi}\Big)
        + 2\ln\Big(\frac{H}{\mu}\Big)
        + \gamma_E
  \bigg]
\nonumber\\
       &&-\,8\epsilon(36-40\epsilon+9\epsilon^2)
  \bigg[
        \psi\Big(\frac{1}{1-\epsilon}\Big)
        +\psi\Big(1-\frac{1}{1-\epsilon}\Big)\bigg]
\nonumber\\
&& \hskip -1cm
 -\,54\epsilon(1\!-\!\epsilon)(2\!-\!\epsilon)
 \bigg[
     \psi\Big(\frac12+\frac{\sqrt{1-14\epsilon+\epsilon^2}}{2(1-\epsilon)}
       \,\Big)
   +\psi\Big(\frac12-\frac{\sqrt{1-14\epsilon+\epsilon^2}}{2(1-\epsilon)}\,
         \Big)
       \bigg]\bigg\}
\nonumber\\
     &+&\mathcal{O}(D-4)
\,,
\label{un-renormalized}
\end{eqnarray}
where $\psi(z)=(d/dz)\Gamma(z)$ denotes the digamma function, and
we made use of
\begin{equation}\label{Hd}
    H^D=\mu^{D-4} H^4\Bigg(1+ ({D-4})\ln\Big(\frac{H}{\mu}\Big)\Bigg)
          +{\cal O}\Big((D\!-\!4)^2\Big)
\,,
\end{equation}
where $\mu$ is an arbitrary renormalization scale.

\bigskip

Since $\epsilon = -\dot H/H^2$ is in general a dynamical quantity,
in order to renormalize the theory properly,
one needs to subtract all divergent terms in Eq.~(\ref{un-renormalized})
containing powers of $\epsilon$. In order to do this we shall make use
of the following counter lagrangian,
\begin{equation}\label{counter}
    \mathcal{L}_c=\sqrt{-g}\bigg(a_0 R^2
     + a_1\kappa g^{\mu\nu}(\partial_\mu\phi)(\partial_\nu\phi)R
     + a_2\kappa g^{\mu\nu}(\partial_\mu\phi)(\partial_\nu\phi)
                 \frac{\partial^2V(\phi)}{\partial^2\phi}
     + a_3[R^2-3R_{\mu\nu}R^{\mu\nu}] \bigg)
\,,
\end{equation}
where the last term denotes the Gauss-Bonnet term in FLRW spaces.
This can be related to the standard form of the Gauss-Bonnet term
by noticing that, since FLRW spaces are conformally flat
and thus have a vanishing Weyl tensor,
$R_{\mu\nu\rho\sigma}R^{\mu\nu\rho\sigma}$ can be expressed as a
linear combination of $R^2$ and $R_{\mu\nu}R^{\mu\nu}$,
\begin{equation}
  R_{\mu\nu\rho\sigma}R^{\mu\nu\rho\sigma}
    = -\frac{2}{(D-1)(D-2)}\left(R^2 - 2(D-1)R_{\mu\nu}R^{\mu\nu}\right)
\,.
\label{Riemann2}
\end{equation}

 In order to fully renormalize the effective
action $\Gamma_{\rm g+gh}$~(\ref{un-renormalized})
nongeometric counterterms are
required. These terms appear as a consequence of including
the terms in the effective action $S_{\rm os}$~(\ref{Sos})
that vanish on shell.
The counter lagrangian~(\ref{counter}) is not unique. Indeed
we could have chosen different counter terms~\cite{JanssenProkopec:2008}.
Since, based on the available information,
there is no unique way to fix the counterterms, the form~(\ref{counter})
of the counter lagrangian suffices for the purpose of this work.

 Varying the individual terms in the counter lagrangian~(\ref{counter})
results in:
\begin{equation}\label{counterterms:1}
    \begin{split}
    \f{1}{V}\f{\delta}{\delta a}\int d^Dx \sqrt{-g} R^2
 &=a^{D-1}H^4\Big(\!-432\epsilon(1-\epsilon)(2-\epsilon)
\\
        &+36\Big(4-34\epsilon+35\epsilon^2-8\epsilon^3\Big)(D\!-\!4)
          +\mathcal{O}\Big((D\!-\!4)^2,\dot \epsilon\Big)
\\
   \f{1}{V}\f{\delta}{\delta a}\int d^Dx
   \kappa\sqrt{-g}g^{\mu\nu}(\partial_\mu\phi)(\partial_\nu\phi)R
  &=a^{D-1}H^4\Big(144\epsilon^2(1-\epsilon)
\\  &    -24\epsilon(2-8\epsilon+5\epsilon^2)(D\!-\!4)\Big)
 +\mathcal{O}\Big((D\!-\!4)^2,\dot \epsilon\Big)
\\
   \f{1}{V}\f{\delta}{\delta a}\int d^Dx\kappa\sqrt{-g}
     g^{\mu\nu}(\partial_\mu\phi)(\partial_\nu\phi)
     \f{\partial^2 V(\phi)}{\partial\phi^2}
       &=a^{D-1}H^4\Big(- 16\epsilon^2(3\!-\!\epsilon)
\\
              &   - 16\epsilon^2(4\!-\!\epsilon)(D\!-\!4)\Big)
 +\mathcal{O}\Big((D\!-\!4)^2,\dot \epsilon\Big)
\\
    \f{1}{V}\f{\delta}{\delta a}\int d^Dx \sqrt{-g}
        [R^2-3R_{\mu\nu}R^{\mu\nu}]
    &= a^{D-1}H^4\Big(36(1-\epsilon)^3\Big)(D-4)
           +\mathcal{O}\Big((D\!-\!4)^2,\dot \epsilon\Big)
\,,
   \end{split}
\end{equation}
where we used~(\ref{curvature invariants}) and in the last step
we again used the background equations of motion~(\ref{background1})
and the following on-shell identities,
\begin{equation}\label{backgrond2}
    \begin{split}
        \sqrt{\kappa}\phi'=\sqrt{2(D-2)\epsilon}aH\;;&
            \qquad  \f{\partial^2 V}{\partial
    \phi^2}(\phi)=2(D-1-\epsilon)\epsilon H^2
    +\mathcal{O}(\dot{\epsilon})
\,.
    \end{split}
\end{equation}
The divergent part of Eq.~(\ref{un-renormalized}) cancels when
the coefficients $a_i$ ($i=0,1,2$)
in the counter lagrangian~(\ref{counter}) are
\begin{equation}
    a_0=-\f{11}{192\pi^2}\f{\mu^{D-4}}{D\!-\!4}+a_0^f
\,,\qquad a_1=\f{13}{288\pi^2}\f{\mu^{D-4}}{D\!-\!4}+a_1^f
\,,\qquad
    a_2=-\f{5}{32\pi^2}\f{\mu^{D-4}}{D\!-\!4}+a_2^f
\,,
\end{equation}
where the $a_i^f$ ($i=0,1,2$) indicates a possible finite part of $a_i$.
$a_3$ remains a free (infinite) parameter.

The renormalized effective action $\Gamma_{\rm 1L\,ren}$ is
then obtained from
\begin{eqnarray}
       \frac{1}{H^{4}a^{D-1}V} \f{\delta\Gamma_{\rm 1L,ren}[a]}{\delta a(l)}
&=&
\frac{1}{16\pi^2}\bigg\{\Big(\beta_0 + \beta_1\epsilon+\beta_2\epsilon^2
                              +\beta_3\epsilon^3+\beta_4\epsilon^4\Big)
\nonumber\\
       &&-\,2\epsilon(198-241\epsilon+63\epsilon^2)
  \bigg[\ln\Big((1-\epsilon)^2\Big)
        + 2\ln\Big(\frac{H}{\bar H_0}\Big)
  \bigg]
\nonumber\\
       &&-\,8\epsilon(36-40\epsilon+9\epsilon^2)
  \bigg[
        \psi\Big(\frac{1}{1-\epsilon}\Big)
        +\psi\Big(1-\frac{1}{1-\epsilon}\Big)\bigg]
\nonumber\\
&& \hskip -1cm
 -\,54\epsilon(1\!-\!\epsilon)(2\!-\!\epsilon)
 \bigg[
     \psi\Big(\frac12+\frac{\sqrt{1-14\epsilon+\epsilon^2}}{2(1-\epsilon)}
       \,\Big)
   +\psi\Big(\frac12-\frac{\sqrt{1-14\epsilon+\epsilon^2}}{2(1-\epsilon)}\,
         \Big)
       \bigg]\bigg\}
\nonumber\\
     &+&\mathcal{O}(D-4)
\,,
\label{renormalized}
\end{eqnarray}
where the coefficients of the terms multiplying
$\epsilon^i/(16\pi^2)$ ($i=0,1,2,3,4$) are given by
\begin{eqnarray}
  \beta_0 &=& -48 + 576\pi^2 (D\!-\!4)a_3
\nonumber\\
  \beta_1 &=& -\frac{2168}{3}
           - 1728\pi^2\Big[8a_0^f + (D\!-\!4)a_3\Big]
       +396\Big[\ln(4\pi)\!-\!\gamma_E + 2 \ln\big({\mu}/{\bar H_0}\big)\Big]
\nonumber\\
  \beta_2 &=& \frac{4352}{3}
           + 192\pi^2\Big[108a_0^f + 12a_1^f - 4a_2^f + 9(D\!-\!4)a_3\Big]
         -482\Big[\ln(4\pi)\!-\!\gamma_E + 2 \ln\big({\mu}/{\bar H_0}\big)\Big]
\nonumber\\
  \beta_3 &=& - \frac{1961}{3}
           + 64\pi^2\Big[-108a_0^f - 36a_1^f + 4a_2^f - 9(D\!-\!4)a_3\Big]
        + 126\Big[\ln(4\pi)\!-\!\gamma_E + 2 \ln\big({\mu}/{\bar H_0}\big)\Big]
\nonumber\\
  \beta_4 &=& 54
\,,
\label{beta:i}
\end{eqnarray}
where $\bar H_0$ is the expansion rate at which
the $\ln(H/H_0)$ term in Eq.~(\ref{renormalized}) vanishes.
The formula~(\ref{renormalized}) is one of the central
results of our work.
Even though $\beta_4$ in Eq.~(\ref{beta:i})
seems to be fully specified by the one loop
calculation, this is in fact not the case. Indeed, one can show that,
upon adding to the counter lagrangian the counterterm
${\cal L}_{c}' = a_4\sqrt{-g} R(\partial^2 V/\partial\phi^2)$,
$\beta_4$ will become a function
of $a_4^f$, and thus unspecified. A similar statement holds for $\beta_0$:
in the absence of the Gauss-Bonnet counterterm $\beta_0$ has a
definite value ($\beta_0 = -48$). Since currently there are
no physical measurements that specify the value
of the Gauss-Bonnet counterterm, we conclude that $a_3$ --
and hence also $\beta_0$ -- is unspecified by the one loop calculation
(see also Refs.~\cite{Koksma:2008jn,JanssenProkopec:2008}).

 Other terms in Eq.~(\ref{renormalized}),
in particular the logarithm and polygamma functions, cannot be
altered by local counterterms, and hence these terms constitute
the physical graviton one loop contributions.
According to the analysis of Ref.~\cite{JanssenProkopec:2008},
when mode mixing is taken account of, the poles of the ghost propagators
coincide with those of the graviton, such that in the full analysis
the digamma functions in the last line of
Eq.~(\ref{renormalized})
are absent (the same holds for the non-renormalized
result~(\ref{un-renormalized})).

\section{Dynamics in quasi de Sitter spaces}
\label{sdynamics}

 Equations~(\ref{effect}) and (\ref{renormalized}),
together with the Bianchi identity, give the quantum modified
Friedmann equations. However, because of the complexity of
(\ref{renormalized}), we will expand the correction in
the limit of small $\epsilon$
(quasi de Sitter space)
\footnote{The question of the dynamics around those $\epsilon$
for which the one-loop potential~(\ref{renormalized})
exhibits poles is addressed in the
companion paper~\cite{JanssenProkopec:2008}.}.

 When expanded in powers of
$\epsilon$ Eq.~(\ref{renormalized}) gives,
\begin{eqnarray}
        \f{1}{a^3V}\f{\delta\Gamma_{1L, \rm ren}[a]}{\delta a(\eta)}
= \frac{H^{4}}{16\pi^2}\Big\{\Big[\beta_0-252\Big]
     +\Big[\beta_1+374-792\ln\Big(\frac{H}{\bar H_0}\Big)+
     792\gamma_E\Big]\epsilon
  +\mathcal{O}(\epsilon^2,\dot\epsilon) \Big\}
 \,.\quad\label{Gamma1L:renormalized:3}
\end{eqnarray}
Inserting (\ref{Gamma1L:renormalized:3}) into Eq.~(\ref{effect})
we obtain the following approximate Friedmann trace equation,
\begin{eqnarray}
&&H^2-\frac{\Lambda}{3}+\frac12\dot H +
  \frac{\beta_0-252}{24\pi}G_N H^4 +
  \frac{33}{\pi}\bigg[\ln\Big(\frac{H}{\bar H_0}\Big)
                   -\gamma_E-\frac{\beta_1+374}{792}
                \bigg]G_N H^2\dot H +{\cal O}(\epsilon^2,\dot\epsilon)
\nonumber\\
&&\hskip 10cm =\, \frac{2\pi G_N}{3}(\rho_M-3p_M)
\,.\qquad
\label{QEOM:trace}
\end{eqnarray}
The quantum correction to the trace of the Einstein equation is
the correction to the expectation value of the trace of
the (quantum) Einstein tensor,
\begin{equation}
\delta G \equiv \langle \Omega|\delta \hat G|\Omega\rangle
  =-\frac{D-2}{2}\langle \Omega|\delta \hat R|\Omega\rangle.
\label{deltaG}
\end{equation}
From the symmetry of the background FLRW space we know that
$\delta G_{\mu\nu}$ contains two independent components:
the first is the trace, and the second can be inferred from the
corresponding Bianchi identity for $\delta G_{\mu\nu}$, which
is a consequence of the Bianchi identity
for the background space Einstein tensor and of
the covariant conservation of the matter stress energy tensor.
Equivalently, one can view $\delta G_{\mu\nu}$
the `stress energy' tensor $(T_{\mu\nu})_Q$
corresponding to the quantum corrections to (\ref{QEOM:trace});
then the symmetries of the FLRW determine its form to be,
\begin{equation}
    (T^\mu{}_\nu)_Q=\text{diag}(\rho_Q,-p_Q,-p_Q,-p_Q)
    \,.
\label{TmnQ}
\end{equation}
The covariant conservation of~(\ref{TmnQ}) implies
the following perfect fluid-like conservation law,
\begin{equation}\label{fluid}
    \frac{d}{dt}(a^4 \rho_Q)=a^4H(\rho_Q-3p_Q)
    \,.
\end{equation}
To solve for $\rho_Q$, we use the following {\it Ansatz:}
\begin{equation}
    \rho_Q=\lambda H^4+\upsilon H^2\dot{H}+(\sigma H^4+\tau
    H^2\dot{H})\ln\Big(\frac{H}{\bar H_0}\Big)
    + \mathcal{O}(\epsilon^2,\dot\epsilon)
\end{equation}
which implies for the fluid equation (\ref{fluid}) that
\begin{equation}\label{fluidtemp1}
    \frac{d}{dt}(a^4\rho_Q)=a^4\Big[4\lambda H^5+(4\upsilon
    +4\lambda+\sigma)H^3\dot{H}+(4\sigma H^5+(4\tau+4\sigma)
    H^3\dot{H})\ln\Big(\frac{H}{\bar H_0}\Big)\Big].
\end{equation}
We read off ($\rho_Q-3p_Q$) from (\ref{QEOM:trace}) and find that
\begin{equation}
    \lambda=-\frac{1}{64\pi^2}(\beta_0-252)\,,
\quad
    \upsilon=\frac{1}{64\pi^2}(\beta_0+\beta_1+122+792\gamma_E)\,,
\quad
     \sigma=0\,,
 \quad\tau=-\frac{99}{8\pi^2}\,,\quad
\end{equation}
and thus
\begin{equation}
    \begin{split}
        \rho_Q&=\frac{1}{64\pi^2}\Big[\!-\!(\beta_0\!-\!252)H^4
        +\Big(\beta_0+\beta_1+122+792\gamma_E\Big)H^2\dot{H}
        -792H^2\dot{H}\ln\Big(\frac{H}{\bar H_0}\Big)\Big]
        +\mathcal{O}(\epsilon^2,\dot\epsilon)\\        
        p_Q&=\frac{1}{64\pi^2}\Big[(\beta_0-252)H^4
        +\Big(\frac13\beta_0-\beta_1-458-792\gamma_E\Big)H^2\dot{H}
        +792H^2\dot{H}\ln\Big(\frac{H}{\bar H_0}\Big)\Big]
        +\mathcal{O}(\epsilon^2,\dot\epsilon)\\
        \rho_Q&+p_Q=\frac{1}{48\pi^2}\big(\beta_0-252\big)H^2\dot{H}
        +\mathcal{O}(\epsilon^2,\dot\epsilon)\,.
    \end{split}
\end{equation}
The quantum corrected Friedmann equations become 
({\it cf.} Eqs.~(\ref{background1})):
\begin{equation}
    \begin{split}
        &3H^2-\Lambda+\frac{\beta_0-252}{8\pi}G_N
        H^4-\frac{\beta_0+\beta_1+122+792\gamma_E}{8\pi}G_NH^2\dot{H}
        +\frac{99}{\pi} G_N
        H^2\dot{H}\ln\Big(\frac{H}{\bar H_0}\Big)
        \\
        &\hskip 4.6cm   
        +\mathcal{O}(\epsilon^2,\dot\epsilon)=8\pi G_N \rho_M
        \\
        &
        -2 \dot{H}-\frac{\beta_0-252}{6\pi}G_N H^2\dot{H}
        +\mathcal{O}(\epsilon^2,\dot\epsilon)
        =8\pi G_N(\rho_M+p_M)
        \,.
    \end{split}
    \label{qdS}
\end{equation}
We shall assume that the matter contribution obeys an equation of
state $p_M=w\rho_M$, with $w$ constant. In this case we can
combine the two equations as
\begin{equation}\label{friedepsklein}
        3H^2+\frac{2\dot{H}}{1+w}-\Lambda+A G_N
        H^4+\bigg[\,B+\frac{99}{\pi}\ln\Big(\frac{H}{\bar H_0}\Big)\,\bigg]
                    G_NH^2\dot{H}
        +\mathcal{O}(\epsilon^2,\dot\epsilon)=0
        \,,
\end{equation}
where we defined
\begin{equation}
    \begin{split}
         A\equiv&\frac{\beta_0-252}{8\pi}\\
        B\equiv&\frac{1}{8\pi}
        \Big[\frac{4}{3(1+w)}
           \Big(\beta_0-252\Big)
        -(\beta_0+\beta_1+122+792\gamma_E)
        \Big]\,.
    \end{split}
\label{AandB}
\end{equation}
Notice that, since we are free to choose $\beta_0$ and $\beta_1$ by a suitable 
choice of the coefficients $a_0^f$ and $a_3$ in the counterterms 
(see Eq.~(\ref{beta:i})) (the Gauss-Bonnet terms must be also included).
Indeed, choosing $\beta_0=252$ and $\beta_1=-374-792\gamma_E$ 
results in $A=0=B$. In fact, $A$ and $B$ are not completely 
independent for a general value of $w$ since from  
(\ref{AandB}) it follows that $B = 4A/[3(1+w)] + {\rm const}$.

\begin{figure}
\begin{center}
\includegraphics[width=3in]{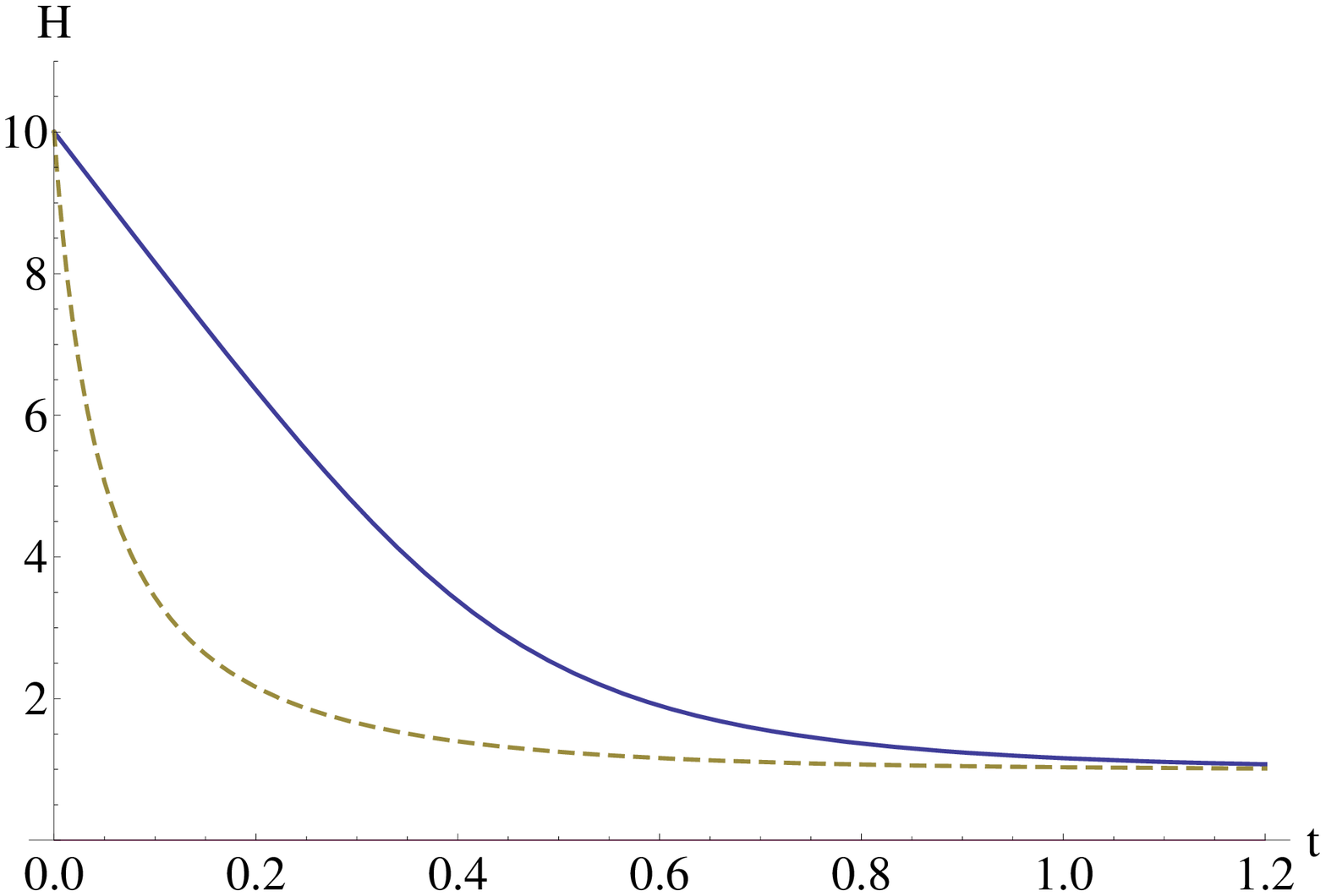}
\includegraphics[width=3in]{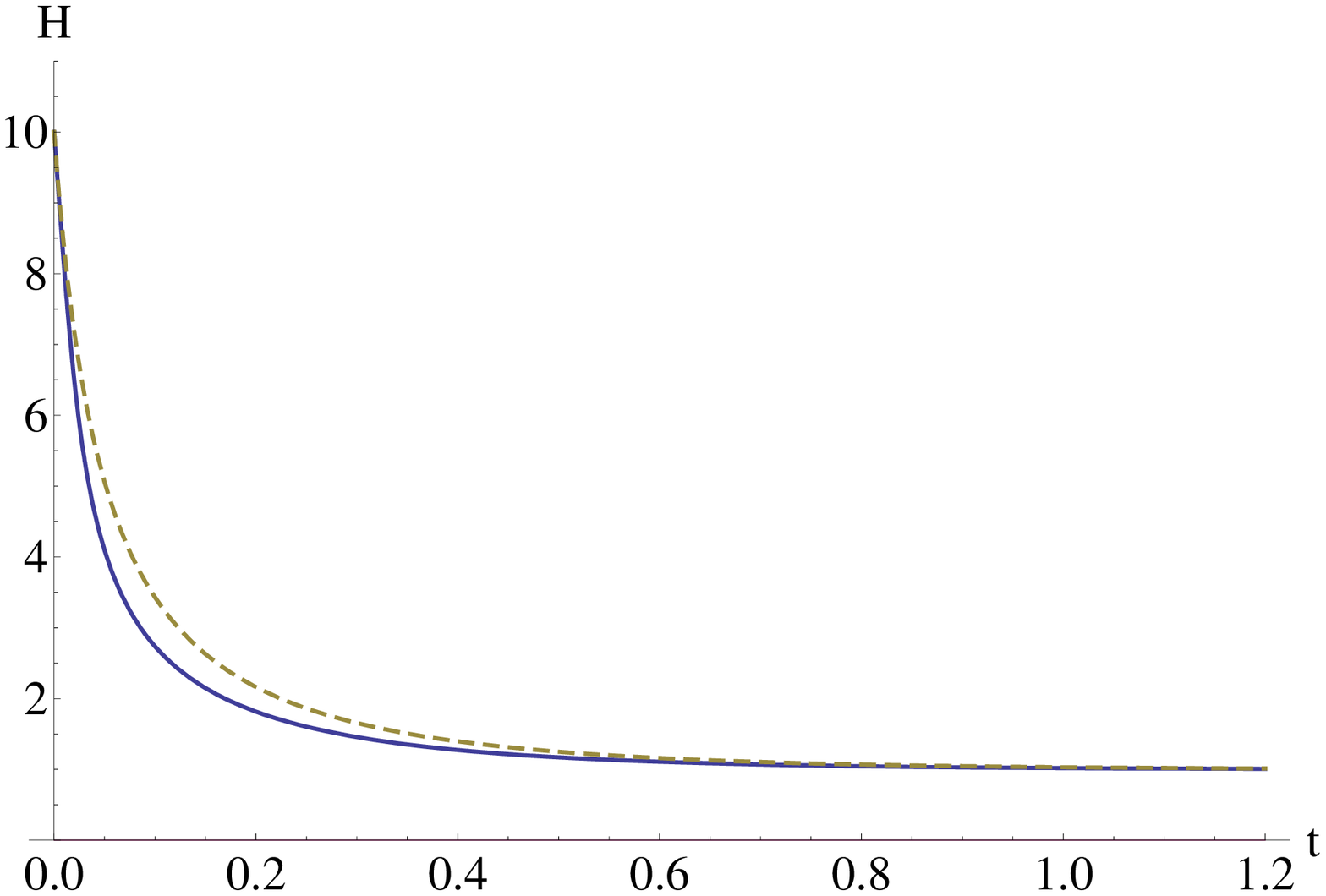}
\caption{Numerical solution to (\ref{friedepsklein}) for $H$ as a
function of time. The {\it red, dashed\/} curve represents the classical
behavior and the {\it blue, solid\/} curve includes our one-loop
corrections. The late time de Sitter limit is clearly obtained.
The quantum corrections lower the effective cosmological constant.
In all plots we choose $\Lambda=3$. This implies that we are
effectively plotting the dimensionless variables:
$h=\sqrt{3/\Lambda}H$; $\tau=\sqrt{\Lambda/3}\,t$ and
$g=(\Lambda/3) G_N$ ($\Lambda=3$ ; $w=1/3$ ; $G_N=0.001$ ;
$A=0$ ; $B=0$ ;  $H(t=0)=10$ ; $\bar H_0=0.1$ (left panel)
$\bar H_0=10$ (right panel)).}\label{hversust}
\end{center}
\end{figure}

 One can integrate Eq.~(\ref{qdS}). The result can be expressed in terms of
 the roots of the quartic equation,
\begin{equation}
    \frac{A}{3}G_NH^4+H^2-\frac{\Lambda}{3} = 0
\,.
\label{Charactersitic eq}
\end{equation}
In the case when $A<0$ all four roots $\pm H_\pm$ are real,
\begin{equation}
 H_\pm^2 = \frac{3}{2AG_N}\left[-1\pm\sqrt{1+\frac{4A}{9}G_N\Lambda}\,\right]
\,.
\label{4roots}
\end{equation}
The positive root $H_+>0$ corresponds to
the one-loop corrected de Sitter attractor.
From equation~(\ref{friedepsklein}) it follows that
$H_+$ is approached exponentially fast. More precisely,
the late time limit can be approximated by the form,
\begin{eqnarray}
 H=H_+\Big[1+2\exp\Big(-\Omega t + \delta_Q\Big)\Big]
 \,,
\label{H:qdS}
\end{eqnarray}
where to order $G_N\Lambda$ and at late times $\Omega t \gg 1$,
\begin{eqnarray}
  \Omega &\simeq& 3(1+w)\bigg\{1 + \frac{G_N\Lambda}{3}
  \bigg[\frac{A}{2}-\frac{1+w}{2}
     \bigg(B+\frac{99}{2\pi}
            \ln\Big(\frac{\Lambda}{3\bar H_0^2}\Big)\bigg)\bigg]\bigg\}
     \sqrt{\frac{\Lambda}{3}}
\label{H:qdS2}
\end{eqnarray}
and $\delta_Q/\Omega$ represents an order $G_N\Lambda$ shift in time, which
is unphysical since it can be absorbed in the definition of time.
This means that quantum effects during quasi de Sitter phase
induce an order $G_N\Lambda$ shift of the late time de Sitter attractor
(which can be read off from $H_+$ in Eq.~(\ref{4roots})).
At late times this de Sitter attractor is approached exponentially fast,
with the characteristic time scale given by $\Omega^{-1}$, which
equals the classical time scale plus an order $G_N\Lambda$ correction,
as expected. In addition, there is an order $G_N\Lambda$ shift $\delta_Q$,
which implies a time delay of $\delta_Q/\Omega$. Note that
$\delta_Q$ can be both positive and negative, depending on the sign and
magnitude of $A$ and $B$ defined in Eq.~(\ref{AandB}).
(The sign of $\delta_Q$ depends also on $\bar H_0$,
but a change in $\bar H_0$ can always be absorbed in a change in $B$.)
This agrees with figure~\ref{hversust}, where we show $H$ as a function
of time both when $\delta_Q$ is positive (left panel)
and when it is negative (right panel) (in the plots we have
chosen $A=0$ and $B=0$).
A positive (negative) correction $\delta_Q$
implies a greater (smaller) expansion rate $H$, and therefore a universe that
has expanded more (less) before entering the late time de Sitter phase.

 At early times the quantum solution
deviates more and more the classical solution,
which approaches the Big Bang singularity at $t=0$.
Formally, the quantum one loop solution is not singular, and
at large and `negative' times (any negative time can be of course
transformed to a positive time by an appropriate time shift)
the solution approaches
the quantum attractor~\cite{Hawking:2002af,Shapiro:2001rh,Koksma:2008jn}
$H\rightarrow H_-$ defined in Eq.~(\ref{4roots}). At
this point the expansion rate becomes of
the order the Planck scale, implying large
higher loop corrections, such that this behavior cannot be trusted.

\begin{figure}
\begin{center}
\includegraphics[width=3in]{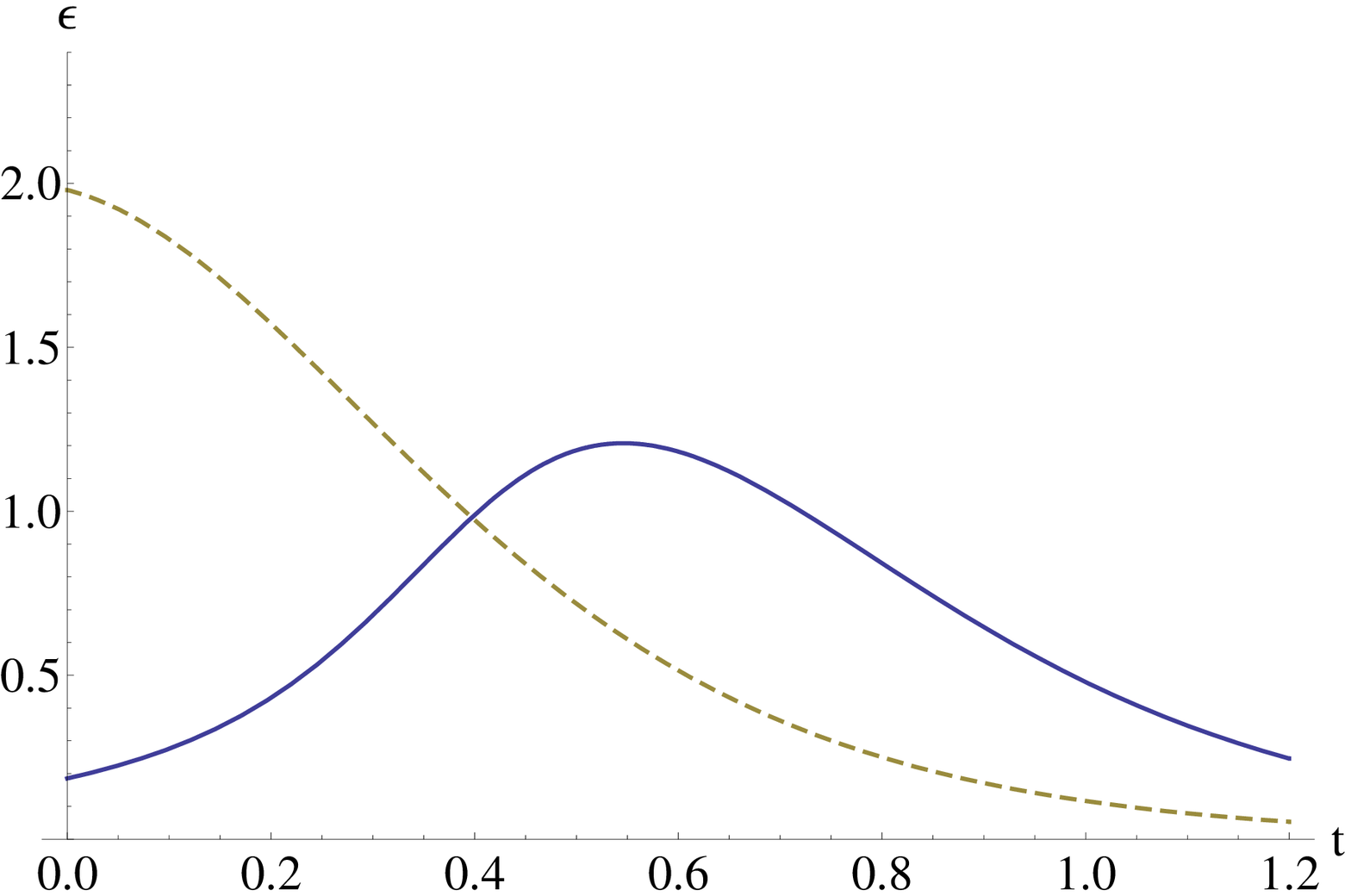}
\includegraphics[width=3in]{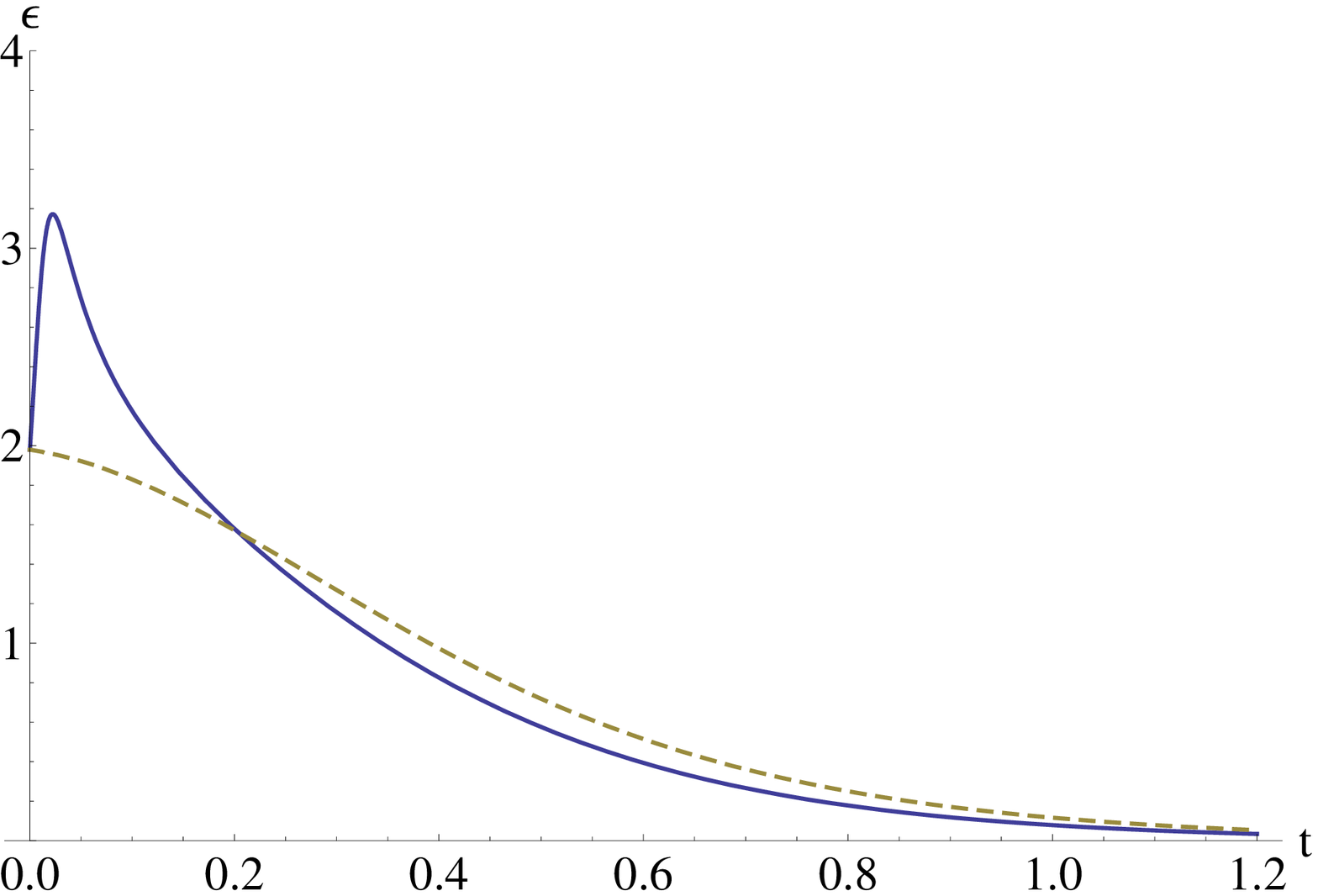}
\caption{$\epsilon$ as a function of time. The {\it red, dashed\/} curve
represents the classical behavior and the {\it blue, solid\/} curve
includes our one-loop corrections. The strong dependence of the
behavior at early times on $\bar H_0$ is clearly visible.
The parameters are: $\Lambda=3$ ; $w=1/3$ ;  $G_N=0.001$ ; $A=0$ ; 
$B=0$ ; $H(t=0)=10$ ; $\bar H_0=0.1$ (left panel) $\bar H_0=10$
(right panel), see also figure~\ref{hversust}.}\label{eversust}
\end{center}
\end{figure}

\section{Discussion}\label{sdisc}

Before specializing the discussion to the two cases discussed
above, we make some general remarks on the validity of our
results. First of all, the correction we calculate is only valid
when $\epsilon$ is strictly constant. A nonconstant $\epsilon$
would induce corrections to the propagators as calculated in
section \ref{sprop} and unfortunately it is not yet known how to
calculate these. This of course does not prevent one from using
these propagators to calculate quantum corrections. One can then
reasonably assume that, as long as in the final answer the change
in $\epsilon$ is sufficiently small, the error one is making is
small and thus the results can be trusted. From figure
\ref{eversust} it is clear however, that there are regimes where
$\epsilon$ is far from constant and one should be careful to trust
our results there.\\
A second general concern is the issue of gauge invariance
(invariance under infinitesimal coordinate transformations). When
both a gravitational field and a matter field are present, the
fluctuations in those fields are coupled and do not transform
independently. Therefore one cannot self consistently quantize the
gravitational fluctuations, without quantizing the matter
fluctuations. The types of structures (e.g. the poles in the
digamma functions), however, are generic since they are naturally
generated by the coincident limit of any propagator of the form
(\ref{scal_prop}), and they do not disappear when matter fluctuations
are included~\cite{JanssenProkopec:2008}.

Because the poles of the digamma functions
$\epsilon_p = 1/2, 2/3, 3/4, .. , 4/3, 3/2$ that yield a divergent
one loop effective potential~(\ref{un-renormalized})
are sufficiently distant from the quasi-de Sitter limit
$\epsilon\rightarrow 0$ considered here, the results of
our dynamical analyisis can be trusted as long as $\epsilon\ll 1/2$.
When this condition is satisfied, quantum effects do not change the fact
that at late times the Universe asymptotes a de Sitter attractor,
albeit with a modified expansion rate given by $H_+$ in Eq.~(\ref{4roots}).
The leading order quantum effect at late times
has a contribution proportional to $H^4$ to the effective energy
momentum tensor. A contribution of this form has also been found
in earlier studies of graviton one-loop effects in de Sitter
space~\cite{Finelli:2004bm}\cite{Tsamis:2005je}\cite{Parker:1969au}.
The exact contribution is unknown because of the ambiguity in
the counterterms. Depending on the choice of counterterms, the
contribution could slightly increase or decrease the effective
late time cosmological constant. Although in our more general
treatment, divergencies appear in the effective action, leading to
the logarithmic correction to (\ref{friedepsklein}),
these corrections have no significant effect at late times.\\
At early times the contribution of quantum effects becomes more
significant. However in this regime we have lost predictability,
since the results strongly depend on the unknown part of the
counterterms and the renormalization scale $\mu$ $(\bar H_0)$.
Moreover, the assumption that $\epsilon<1/2$ and nearly constant
appears to be violated.

\section{Conclusion and outlook}\label{sconclusion}

In this paper we calculated the quantum corrected Friedmann
equations due to the one loop vacuum bubble from gravitons in a
FLRW universe with constant $\epsilon\equiv-{\dot{H}}/{H^2}$.
The result has a divergence that contains terms proportional to 
$\epsilon H^4$, $\epsilon^2 H^4$ and $\epsilon^3 H^4$, which can be renormalized
using local counterterms, which include both geometric and scalar field
counterterms. This is consistent with the
result that in de Sitter space ($\epsilon=0$) one loop effects
lead to a finite constant shift of the cosmological term $\propto H^4$.
We study the dynamics in the quasi de
Sitter limit and find that they are not much different from the
dynamics in true de Sitter space. Indeed,
the quantum effects induce a shift in
the effective, late time, cosmological constant
$\sim ({A}/{36})G_N H^4$,
where $A$ is an unknown parameter, that can be expressed in terms of
the Gauss-Bonnet counterterm  with an ${\cal O}(1/(D-4))$ coefficient.

 Although our results are correct
within the approximations used, the results described above should not
be taken too literally. The propagators we used (and hence the
singularity structure we find) are strictly speaking only valid
when $\epsilon={\rm constant}$. Our analysis is correct as long
as any time variation in $\epsilon$ is small enough, which is indeed
the case sufficiently close to de Sitter space. Indeed,
our late time solution does have $\dot{\epsilon} \rightarrow 0$.
Therefore we have good reasons to
believe that our solution approximates well the solution
of the full theory, at least at late times and sufficiently
close to de Sitter space.

 Another issue is that we choose our propagator such to describe a
physically meaningful vacuum state. However, due to the evolution
and mixing of modes, close to the de Sitter
attractor the Universe will not be in a vacuum state,
but in some excited state (that can be described by mode mixing
in momentum space), which might influence our results.
We postpone a study of this question for future publication.

 The next issue is the question of gauge invariance.
Since there is both matter and gravity in our model,
one should self-consistently
take both fluctuations in matter and gravitons into account. This
issue is complicated due to the mixing of the degrees of freedom,
and hence it is addressed in a separate
publication~\cite{JanssenProkopec:2008}.
Taking this mixing into account changes our results
quantitatively, but since the singularity structure is inherent in
the propagators (and those do not change), the logarithmic terms do not
cancel, such that qualitative features
of the analysis presented here remain unchanged.

 Finally, an important question is what are the dynamics near the poles
of the digamma functions $\epsilon_p = 1/2, 2/3, 3/4, .. , 4/3, 3/2$
where the one loop effective potential~(\ref{renormalized})
diverges. The corresponding dynamical analysis is performed in
Ref.~\cite{JanssenProkopec:2008}. Here we just note that the Universe
typically gets stuck near the poles, such that each pole acts as
a late time attractor. Probably the most important attractors are
the two highest poles $\epsilon_p = 3/2$ and $\epsilon_p = 4/3$ 
(the latter is also the value of $\epsilon$ in matter era).
The latter pole is the late time attractor of a universe filled mostly
with radiation and a cosmological term~\cite{JanssenProkopec:2008}, 
which represents a realistic
composition of the Universe immediately after the Big Bang.

\section*{Acknowledgements}

We would like to thank Richard P. Woodard for useful discussions.
The authors acknowledge financial support by Utrecht University.
T.P. acknowledges financial support by FOM grant 07PR2522.

\end{document}